\documentclass[useAMS,usenatbib]{mn2e}
\usepackage{amssymb}
\usepackage{graphicx}
\usepackage{xspace}

\usepackage{ifthen}
\def\draftversion{1} % set this to 1 to display the content of the draft commands, or to 0 to hide them.

\ifthenelse{\equal{\draftversion}{0}}{
	\usepackage{xcolor}
	\newcommand{\tmp}{}
	\newenvironment{envcomm}[1]{\renewcommand{\tmp}{#1}\begin{color}{blue}\begin{center}\hrule\vspace{0.5mm}\tmp's COMMENTS\end{center}}{\begin{center}END OF \tmp's COMMENTS\vspace{0.5mm}\hrule\end{center}\end{color}}
	\newenvironment{draft}{\begin{color}[rgb]{0,0.4,0}\begin{center}\hrule\vspace{0.5mm}DRAFT\end{center}}{\begin{center}END OF DRAFT\vspace{0.5mm}\hrule\end{center}\end{color}}
	\newcommand{\comcomm}[2]{\begin{color}{blue}\ $\bullet$ \textbf{#1:} #2 $\bullet$\ \end{color}}
	\newcommand{\revend}[1]{\par\begin{color}[rgb]{0,0.4,0}\begin{center}\hrule\vspace{0.5mm}END OF #1's REVISIONS\vspace{0.5mm}\hrule\end{center}\end{color}\par}
	\newcommand{\todo}[1]{\begin{color}{red}\ $\bullet$ \textbf{To do: }#1 $\bullet$\ \end{color}}
	\newcommand{\new}[1]{\begin{color}{red}#1\end{color}} % real size
	\newcommand{\del}[1]{\begin{color}[rgb]{0,0.5,0.0}\ $\bullet$ \textbf{Deleted: }#1 $\bullet$\ \end{color}}
	\newcommand{\sk}[1]{\begin{color}[rgb]{0.6,0,0.6}#1\end{color}}
	}{
	\newsavebox{\trashcan}
	\newenvironment{envcomm}[1]{\begin{lrbox}{\trashcan}\begin{minipage}{\columnwidth}}{\end{minipage}\end{lrbox}}
	
	\newcommand{\comcomm}[2]{}
	\newcommand{\revend}[1]{}
	\newcommand{\todo}[1]{}
	\newcommand{\new}[1]{#1}
	\newcommand{\del}[1]{}
	\newcommand{\sk}[1]{}
	}
\newcommand{\aj}{AJ}% Astronomical Journal
\newcommand{\araa}{ARA\&A}% Annual Review of Astron and Astrophys
\newcommand{\apj}{ApJ}% Astrophysical Journal
\newcommand{\apjl}{ApJ}% Astrophysical Journal, Letters
\newcommand{\apjs}{ApJS}% Astrophysical Journal, Supplement
\newcommand{\aap}{A\&A}% Astronomy and Astrophysics
\newcommand{\aapr}{A\&A~Rev.}% Astronomy and Astrophysics Reviews
\newcommand{\mnras}{MNRAS}% Monthly Notices of the RAS
\newcommand{\mh}{\ensuremath{\textrm{\,--\,}}}
\newcommand{\bb}[1]{\ifmmode \mbox{\boldmath $ #1$} \else  \mbox{\boldmath $#1$} \fi}

\newcommand{\U}[1]{\ensuremath{\mathrm{~#1}}}

\newcommand{\Myr}{\U{Myr}}

\newcommand{\pc}{\U{pc}}
\newcommand{\kpc}{\U{kpc}}

\newcommand{\msun}{\U{M}_{\odot}}
\newcommand{\Msun}{\msun}
\newcommand{\cc}{\U{cm^{-3}}}
\newcommand{\kms}{\U{km\ s^{-1}}}

\newcommand{\mach}{\ensuremath{\mathcal{M}}}

\newcommand{\ramses}{{\tt RAMSES}\xspace}

\newcommand{\fig}[2][]{Figure#1~\ref{fig:#2}}

\renewcommand{\fig}[2][]{Fig#1.~\ref{fig:#2}}

\newcommand{\citetip}[1]{#1 (in prep.)}
\newcommand{\citepip}[1]{(#1, in prep.)}

\title[Starbursts from tides and turbulence]{Starbursts triggered by inter-galactic tides and\\interstellar compressive turbulence}

\author[Renaud, Bournaud, Kraljic \& Duc]{Florent~Renaud\thanks{florent.renaud@cea.fr}, Fr\'ed\'eric~Bournaud, Katarina Kraljic \& Pierre-Alain Duc\\
Laboratoire AIM Paris-Saclay, CEA/IRFU/SAp, CNRS, Universit\'e Paris Diderot, F-91191 Gif-sur-Yvette Cedex, France}

\date{Accepted 2014 March 27. Received 2014 March 26; in original form 2014 March 10}

\begin{document}
\maketitle

\newcommand{\pdfwidth}{\delta}
\newcommand{\Bournaud}{Bournaud et al.}
\newcommand{\Duc}{Duc, Bournaud \& Renaud}
\newcommand{\Gieles}{Gieles \& Renaud}
\newcommand{\Renaud}{Renaud, Bournaud \& Duc}

%%%%%%%%%%%%%%%%%%%%%%%%%%%%%%%%%%%%%%%%%%%%%%%%%%%%%%%%%%%%%%%%%%%%%%%%%%%%%%%%
\begin{abstract}
Using parsec-resolution simulations of a typical galaxy merger, we study the triggering of starbursts by connecting the (inter-)galactic dynamics to the structure of the interstellar medium. The gravitational encounter between two galaxies enhances tidal compression over large volumes, which increases and modifies the turbulence, in particular its compressive mode with respect to the solenoidal one. This generates an excess of dense gas leading to intense star formation activity. Along the interaction, the compressive turbulence modifies the efficiency of gas-to-star conversion which, in the Schmidt-Kennicutt diagram, drives the galaxies from the sequence of discs to that of starbursts.
\end{abstract}

%%%%%%%%%%%%%%%%%%%%%%%%%%%%%%%%%%%%%%%%%%%%%%%%%%%%%%%%%%%%%%%%%%%%%%%%%%%%%%%%
\section{Introduction}

Starbursts are generally attributed to galaxy interactions and mergers \citep[e.g.][]{Sanders1996}. Mergers induce gravitational torques and global gas inflows toward the galaxy centres, enhancing the gas surface density \citep{Keel1985}. Simulations have demonstrated that this process triggers bursts of star formation, especially in the inner regions of advanced mergers \citep[e.g.][]{Barnes1991, Hopkins2006, Robertson2006, DiMatteo2007, Cox2008, Karl2010}. Yet, this process alone does not explain all properties of merger-induced star formation, which can be intense even in early interaction phases \citep{Ellison2008}, and is often spatially-extended, off-nuclear \citep{Wang2004, Barnes2004, Cullen2006, Elmegreen2006a, Smith2008, Hancock2009, Chien2010}. In fact, starburst galaxies can convert their gas into stars an order of magnitude faster than isolated discs with similar global gas surface densities, i.e. independently of the global compression by inflows \citep{Daddi2010b, Genzel2010, Saintonge2012}. 

Simulations have shown that, on top of the global enhancement of the gas surface density by the interaction-induced inflows, changes in the sub-structure of the interstellar medium (ISM) can control the starburst activity of mergers \citep{Teyssier2010, Powell2013}. Interactions increase the ISM turbulence, in agreement with observations \citep{Irwin1994,Elmegreen1995b}\new{, and could help compress the diffuse gas reservoirs \citep{Jog1992}}. Nevertheless, it remains unknown (i) whether numerical models can now reach convergence on the global starburst activity by sufficiently resolving the small-scale physics and structure of the ISM, (ii) through which physical processes galaxy mergers could increase or modify turbulence, and (iii) why triggered turbulence would lead to starburst activity rather than stabilizing clouds against collapse.

Using parsec-scale simulations of a representative merger, we here address these three aspects by probing the role of the tidal field in modifying the properties of the ISM turbulence down to the scales of star-forming regions.

%%%%%%%%%%%%%%%%%%%%%%%%%%%%%%%%%%%%%%%%%%%%%%%%%%%%%%%%%%%%%%%%%%%%%%%%%%%%%%%%
\section{Parsec-scale ISM structure in a prototypical merger simulation}
\label{sec:simu}

We use a hydrodynamical simulation of a major galaxy merger run with the adaptive mesh refinement code \ramses \citep{Teyssier2002}, and presented in detail in \citetip{\Renaud}. \new{The physical ingredients implemented are identical to those in \citet{Bournaud2014}. They include heating and cooling at solar metallicity (down to $100 \U{K}$)}, star formation and stellar feedback (photo-ionization, radiative pressure and supernovae thermal blasts). The refinement is based on the density of baryons and ensures that the Jeans length is always resolved by at least four cells. The resolution reached is $1.5 \pc$, with a gas mass per cell of $\approx 1000 \Msun$. The galaxy models are made of live dark matter and stellar components on top of the gas discs. 

Although the merger model ressembles the Antennae galaxies (NGC~4038/39) at some specific time, its parameters are common for $\Lambda$CDM merger orbits (elliptical orbit with an impact parameter of the order of twice the galactic radius, and with a relative velocity of $160 \kms$, see \citealt{Khochfar2006}).

We also run the same merger model at lower resolutions (6 and $24 \pc$), and the progenitor galaxies in isolation, as a control sample. The numerical methods used to compute the tidal field and the turbulent modes are given in Appendix~\ref{app:computation}.

The thumbnails on \fig{evol} show the density maps of the gas component at a few stages of the merger. The two zoom-in views reveal the fragmentation of the ISM at an early stage of the merger, before galaxy-scale inflows concentrate the gas toward the nuclei. The bottom panel of \fig{evol} shows the star formation history of the merger. Data from the low resolution run ($24 \pc$) shows variations of a factor of a few compared to the other runs (1.5 and $6 \pc$) for which convergence is reached. This demonstrates that parsec-resolution simulations allow to capture the small-scale physics relevant for the global star formation activity.

\begin{figure*}
\includegraphics[width=\textwidth]{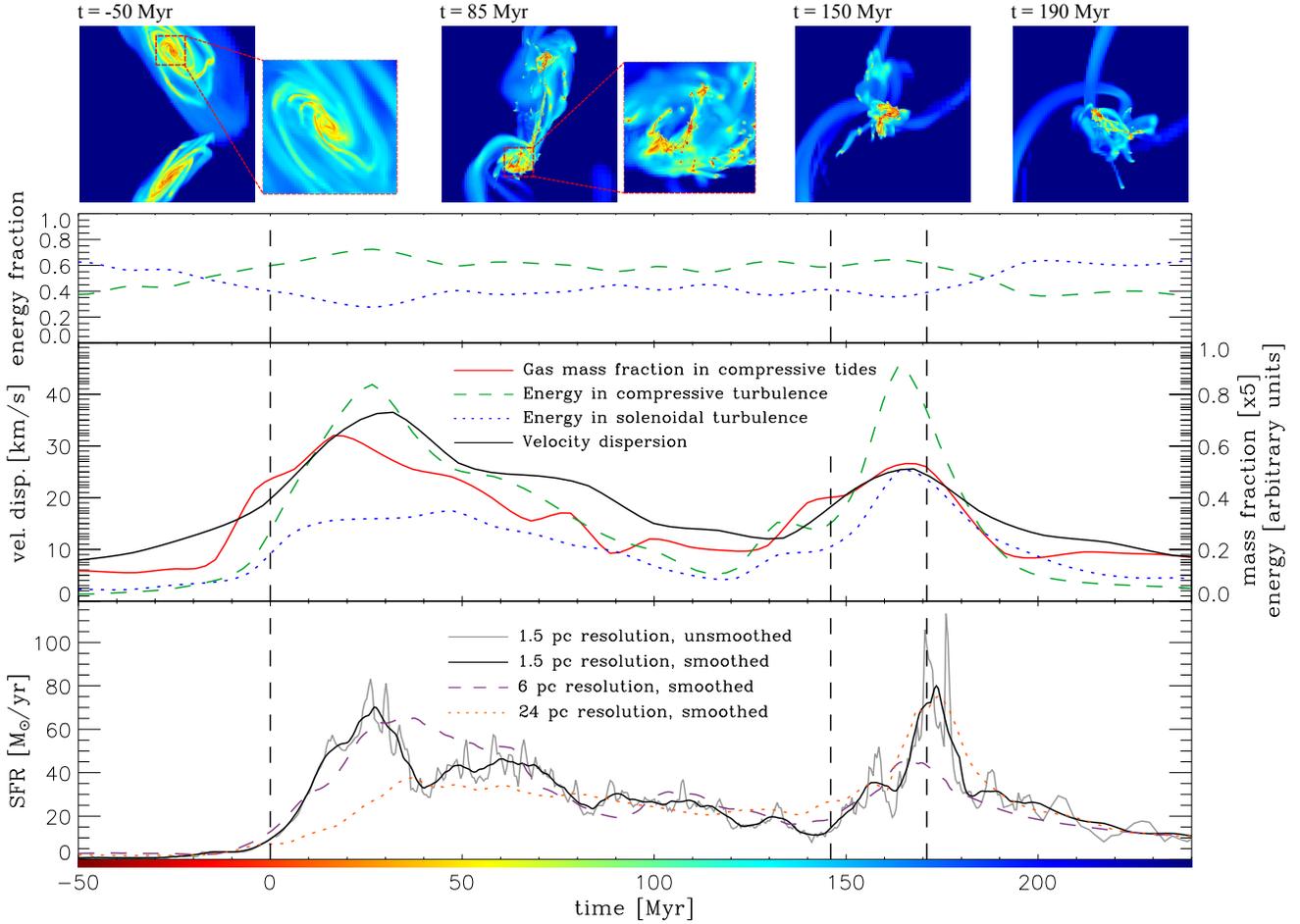}
\caption{Thumbnails: Density maps of gas at four stages of the merger in the central $30 \kpc \times 30 \kpc$, and two zoom-in into the central region ($5 \kpc \times 5 \kpc$) of one of the galaxies, revealing the fragmentation of the ISM already induced by the first encounter. Top: energy fraction of the compressive and solenoidal turbulent modes to the total turbulent energy. Middle: gas mass fraction in fully compressive tides, energy in compressive and solenoidal turbulent modes, and one-dimensional velocity dispersion. Bottom: star formation rate at different resolutions (1.5, 6 and $24 \pc$). The curves have been smoothed for the sake of clarity, but we provide an unsmoothed version, for reference. The vertical dashed lines indicate the three galactic pericenter passages: the progenitors are in isolation for $t \lesssim 0$, and the final coalescence begins at $t\approx 190 \Myr$. The colour bar indicates the coding of the time used in the other figures of this Letter.}
\label{fig:evol}
\end{figure*}

%%%%%%%%%%%
\subsection{Compressive tides}
\label{sec:tides}

A general property of galaxy interactions is that they are prone to extended regions of compressive tides\footnote{The tidal field experienced by a spatially extended object is fully compressive (or ``compressive'' for short) when the tidal forces are pointing inward along all axes, as opposed to the extensive mode in which forces point outward along at least one axis (see e.g. \citealt{Valluri1993, Dekel2003}).} \citep{Renaud2009} while isolated discs may host compressive tides only in very narrow, cored regions \citep{Emsellem2008}. \citet{Renaud2008} noted a correspondance between the position, duration and energy of the compressive tides and the observed properties of young star clusters in local mergers, and suggested that the compressive nature of tides could help to trigger the formation of stars. Recently, \citet{Jog2013a,Jog2013b} provided an analytical argument supporting this idea by deriving the modified Jeans and Toomre stability criteria in the presence of tides and showed that, indeed, compressive tides favor the collapse of gaseous structures. 

Following \citet{Renaud2008}, we show in \fig{evol} that the gas mass fraction in compressive tidal mode increases by a factor $\approx 5$ during the galaxy pericenter passages. Note that this quantity starts to increase even before the first pericenter passage itself ($t=0$) because of the long-range nature of the gravitational effect of tides. At final coalescence ($t > 190 \Myr$), the mass fraction in compressive tides is almost back to its initial value of a few percents. Such evolution is quantitatively representative of a wide variety of mergers, as shown in \citet{Renaud2009}.

%%%%%%%%%%%
\subsection{Compressive and solenoidal turbulence}
\label{sec:turbulence}

Turbulent motions can be decomposed into compressive (curl-free) and solenoidal (divergence-free) modes \citep{Kritsuk2007, Federrath2010}. In the inertial range (i.e. the scales between the injection of turbulence and its dissipation, \citealt{Frisch1995}), equipartition is reached when the compressive mode carries 1/3 of the turbulent energy and the solenoidal component carries the other 2/3 \citep{Federrath2010}\footnote{This is because compression acts along one dimension while solenoidal mixing is a two-dimensional process.}, which is about the case in isolated and pre-merger galaxies in our simulation (see \fig{evol} for $t \lesssim -20 \Myr$). Such turbulence can be injected by feedback, gravity and hydrodynamics \citep{Elmegreen2004, MacLow2004, Hennebelle2012}.

When a galaxy collision occurs, the gas velocity dispersion increases by a factor of a few (from $8 \kms$ to $36 \kms$ in our simulation, see \fig{evol}), in line with the findings of \citet{Irwin1994}, \citet{Elmegreen1995b}, \citet{Bournaud2011b} and \citet{Ueda2012}. We measure that the energy in compressive mode rises by a factor $\approx 12$, while that in solenoidal mode only increases by a factor of $\approx 5$, such that equipartition is no longer sustained. The rise of the turbulent mode energies shortly follows that of the mass fraction in compressive tides, with a delay of $\approx 10 \Myr$. The two effects concur throughout the merger, suggesting the tidal one excites the turbulent one.

%%%%%%%%%%%%
\subsection{Gas density PDF}
\label{sec:pdf}

\begin{figure}
\includegraphics[width=\columnwidth]{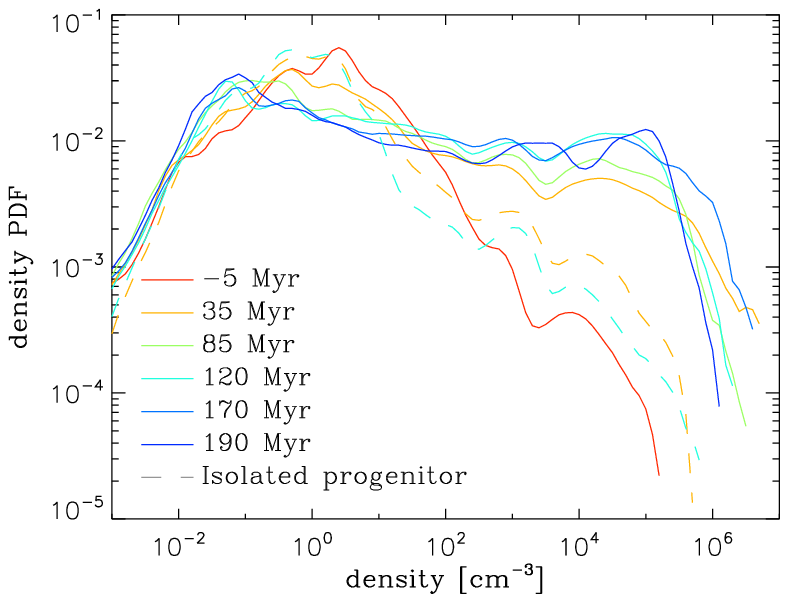}
\caption{Normalized mass-weighted probability distribution function of the gas, at several epochs for the merger (solid lines), and for a progenitor galaxy run in isolation (dashed lines).}
\label{fig:pdf}
\end{figure}

The modification of the ISM turbulence affects the gas density probability distribution function (PDF), as shown in \fig{pdf} at several stages along the merger. In isolation, the PDF can be, at first order, approximated by a log-normal functional form. Theoretically, such log-normal shape is characteristic of isothermal supersonic medium \citep{Vazques1994, Nordlund1999, Wada2001}, but it provides a good fit to the PDF in many galaxy simulations \citep[e.g.][]{Tasker2008, Robertson2008, Bournaud2011b}. On top of the log-normal, a power-law tail at densities $\gtrsim 10^3 \cc$ is captured in high-resolution simulations. It corresponds to the self-gravitating gas \citep{Elmegreen2011} and represents about two percent of the gas mass in our isolated discs. 

To first order, the width $\pdfwidth$ of the PDF is connected to the Mach number $\mach$ through $\pdfwidth^2 = \ln{\left[1+(1-2\zeta/3)^2\mach^2\right]}$, where $\zeta = 0$ (resp. 1) represents a purely compressive (resp. solenoidal) turbulent mode, and equipartition corresponds to $\zeta = 1/2$ \citep{Federrath2008, Molina2012}. After the first encounter, the maximum density increases significantly, as a natural result of the raise of the turbulent energy noted above\footnote{The Mach number depends on both the velocity dispersion and the temperature. While the former increases significantly during the merger, the latter remains almost constant.}. Since the compressive mode overcomes the solenoidal one ($\zeta < 1/2$), the width $\pdfwidth$ increases faster with $\mach$ than it would at equipartition, which results in a secondary component of the PDF at high density. This additional component contains between 10 and 20 percent of the gas mass, depending on the stage of the merger, i.e. much more than the initial power-law tail (two percent). 

%%%%%%%%%%%%
\subsection{Star formation}
\label{sec:sf}

The star formation rate (SFR) is known to be mostly tightly correlated with the amounts of high-density gas \citep[$\gtrsim 10^4 \cc$,][\new{but see \citealt{Longmore2013} in specific small-scale environments}]{Gao2004, Garcia2012}. The large excess of dense gas induced by the triggered compressive turbulence should thus lead to an intense starburst independently of any theoretical model for the small-scale SFR. In our simulations, we quantify this effect with a standard recipe in which the local SFR density depends on the gas mass density divided by the gravitational free-fall time (for the gas denser than $50 \cc$, and with a two percent efficiency, \citealt{Krumholz2007a}).

The global SFR yields the expected bursts at about the times of the pericenter passages (\fig{evol}). \new{In the end, the successive increases of the compressive tides, the compressive turbulence and the SFR occur with only $10\mh 30 \Myr$ between the initial trigger (tides) and the resulting starburst. This small delay confirms that the increase of velocity dispersion in mergers is not a consequence of stellar feedback but is instead of gravitational origin \citep[as already stated by][]{Teyssier2010, Powell2013}.}

\begin{figure}
\includegraphics[width=\columnwidth]{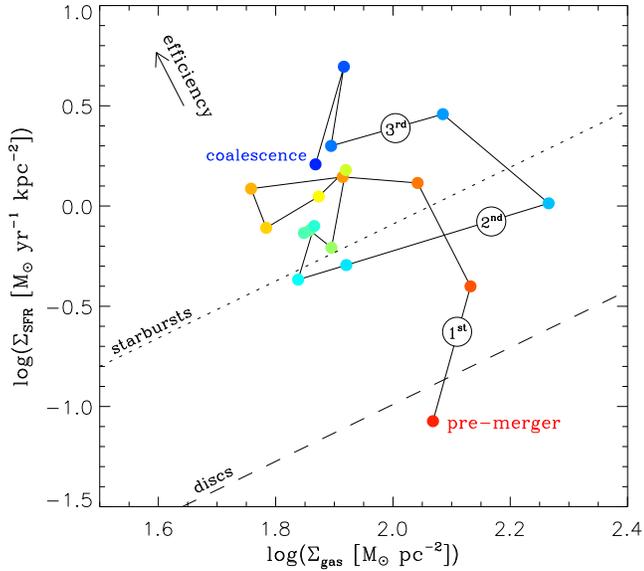}
\caption{Evolution of surface density of gas and of star formation rate inside the half-mass radii of the galaxy(ies) ($2 \pm 0.3 \kpc$), every $\approx 10 \Myr$ along the merger. Colour running from red to blue codes time, as in the colour bar of \fig{evol}. The numbers in circles indicate the three encounters ($t = 0$, $146 \Myr$ and $171 \Myr$). The dashed and dotted lines indicate the sequences of discs and of starbursts, as in \citet{Daddi2010b}.}
\label{fig:ks}
\end{figure}

\fig{ks} shows the evolution of the galactic system in the Schmidt-Kennicutt diagram. The pre-merger galaxies and the isolated progenitors are initially on the sequence of discs, i.e. forming stars with the same efficiency as local spirals or high redshift discs. During the first starburst episode, the increase of compressive turbulence makes their ISM more efficient at converting gas into stars, although global inflows did not have time to significantly enhance the surface density of gas yet. The system moves to the starburst regime. At the second encounter, the surface density of gas gets boosted by gravitational torques and inflows on top of the compressive triggers already seen at the first encounter. This translates into a new enhancement of the star formation efficiency.

%%%%%%%%%%%%%%%%%%%%%%%%%%%%%%%%%%%%%%%%%%%%%%%%%%%%%%%%%%%%%%%%%%%%%%%%%%%%%%%%
\section{Discussion and conclusion}
\label{sec:summary}

Using hydrodynamical simulations, we propose a physical explanation for the enhancement of star formation activity in galaxy mergers. Our main findings are as follows:
\begin{itemize}
\item The global SFR evolution reaches numerical convergence at parsec-scale resolution.
\item The rise of the gas mass fraction in compressive tides in extended volumes during the galactic collisions pumps turbulence into the ISM and unbalances the equipartition between compressive and solenoidal turbulence modes. \new{Such turbulence is not primarily driven by feedback.}
\item The compressive turbulence allows to overcome the regulating, stabilizing effect of turbulence, and to generate an excess of dense gas.
\item This excess translates into an enhanced star formation activity and drives the merger to the starburst regime in the Schmidt-Kennicutt diagram. 
\item \new{From the gravitational and tidal trigger to the ignition of starburst, the full sequence takes $\sim 10\mh 30 \Myr$.}
\end{itemize}

Here, we have only accounted for \emph{integrated} properties of the tides, turbulence and star formation. A study of the spatial distribution in the galaxies as well as the propagation of these phenomena will be described in a forthcoming contribution \citepip{\Bournaud}, showing in particular that the locations in space and time of the compressive tides, compressive turbulence and star forming regions coincide.

We have drawn our conclusions using a simulation of the Antennae galaxies. Obviously, in other interacting systems, the quantitative results we presented here are modulated by the parameters of the galaxies (shape of the halo, mass ratio, etc.) and by the details of their interaction (spin-orbit coupling, impact parameter, orbital eccentricity, etc.). However, the ubiquity of compressive tides in mergers has been previously demonstrated by \citet{Renaud2009}, which suggests that the triggers and physical processes mentioned above exist in many mergers.

An increase of the turbulent Mach number maintaining equipartition between compressive and solenoidal modes would widen the PDF without changing its functional form (from a log-normal), and would drive the evolution of the progenitor galaxies at constant efficiency in the Schmidt-Kennicutt diagram. This explains the similar star formation efficiencies between local spirals and redshift-two discs which have turbulent speeds about ten times larger \citep[see also \citealt{Kraljic2014}]{Renaud2012}. A second component at high density in the PDF is necessary to reach the regime of local starbursts (i.e. with a SFR comparable to that of redshift-two discs but with a less turbulent ISM). This is generated by the deviation from equipartition which occurs during a merger. \new{\citet{Krumholz2012} neglected the second component of the PDF in their conversion of volume to surface densities. By doing so, the starbursting mergers lie on the same regime as disc galaxies in their model, which leads to an apparent ``universality'' of star formation.}

The physical context of mergers naturally modifies the turbulence forcing in ways that have so far been only arbitrarily implemented in idealised volume-limited simulations of the ISM \citep{Federrath2008, Seifried2011, Saury2014}. We have shown here that the (10)kpc-scale forcing induces variations in the structure of the ISM at parsec-scale, but it remains unclear whether the turbulence would retain a signature of these variations at the much smaller scale of dissipation. If it does, this might result in a variation of the proto-stellar core mass function, as proposed by \citet{Hopkins2012b}, and of the stellar initial mass function in mergers, with respect to isolated galaxies.

%%%%%%%%%%%%%%%%%%%%%%%%%%%%%%%%%%%%%%%%%%%%%%%%%%
\section*{Acknowledgments}
We thank Patrick Hennebelle, Christoph Federrath and Changa Jog for stimulating discussions, and the referee Diederik Kruijssen for insightful comments. This work was supported by GENCI and PRACE resources (2013,2014-GEN-2192 and pr86di allocations) and by the European Research Council through grant ERC-StG-257720.

%%%%%%%%%%%%%%%%%%%%%%%%%%%%%%%%%%%%%%%%%%%%%%%%%%%%%%%%%%%%%%%%%%%%%%%%%%%%%%%%
\appendix
\section{Computation of the tides and the turbulence modes}
\label{app:computation}

The tidal field is computed with a first order finite difference scheme over a scale of $40 \pc$, using the total gravitational force provided by \ramses. This way, we obtain the components $T_{ij} = -\partial_i \partial_j \phi$ of the tidal tensor for the potential $\phi$. The eigenvalues of the tensor indicate the strength of the tidal forces. The tides are (fully) compressive when all eigenvalues are negative.

We measure compressive and solenoidal turbulent motions by splitting the simulation volume into sub-volumes of $2^3$ cells of $\epsilon = 40 \pc$ each. In each sub-volume, the average motion is canceled out to keep only the local dispersion field $\bb{v}$. We re-construct the compressive turbulent mode as $v_{\textrm{comp},i} = \epsilon \partial_i v_i$ so that it has the same divergence as the original field ($\nabla \cdot \bb{v}$) and no curl. The solenoidal mode $v_{\textrm{sol},i}= v_i - v_{\textrm{comp},i}$ recovers the curl of the local dispersion field ($\bb{\nabla} \times \bb{v}$) with zero divergence. The turbulent energy estimated this way, summed up over the whole volume, recovers that obtained from the evolution of the mass-weighted total velocity dispersion (\fig{evol}). \new{\citet{Federrath2011} noted that the solenoidal mode could be underestimated if computed over a few resolution elements. By evaluating it at the $40 \pc$ scale, and since equipartition is reached when our galaxies are isolated, we do not suffer from such bias.} The turbulent cascades from scales larger than $40 \pc$ down to the parsec-scale will be detailed in \citetip{\Bournaud}.

\bibliographystyle{mn2e}

\small
\begin{thebibliography}{}

\bibitem[\protect\citeauthoryear{{Barnes}}{{Barnes}}{2004}]{Barnes2004}
{Barnes} J.~E.,  2004, \mnras, 350, 798

\bibitem[\protect\citeauthoryear{{Barnes} \& {Hernquist}}{{Barnes} \&
  {Hernquist}}{1991}]{Barnes1991}
{Barnes} J.~E.,  {Hernquist} L.~E.,  1991, \apjl, 370, L65

\bibitem[\protect\citeauthoryear{{Bournaud}, {Chapon}, {Teyssier}, {Powell},
  {Elmegreen}, {Elmegreen}, {Duc}, {Contini} \& {et al.}}{{Bournaud}
  et~al.}{2011}]{Bournaud2011b}
{Bournaud} F.,  {Chapon} D.,  {Teyssier} R.,  {Powell} L.~C.,  {Elmegreen}
  B.~G.,  {Elmegreen} D.~M.,  {Duc} P.-A.,  {Contini} T.,    {et al.} 2011,
  \apj, 730, 4

\bibitem[\protect\citeauthoryear{{Bournaud}, {Perret}, {Renaud}, {Dekel},
  {Elmegreen}, {Elmegreen}, {Teyssier}, {Amram}, {Daddi}, {Duc}, {Elbaz},
  {Epinat}, {Gabor}, {Juneau}, {Kraljic} \& {Le Floch'}}{{Bournaud}
  et~al.}{2014}]{Bournaud2014}
{Bournaud} F.,  {Perret} V.,  {Renaud} F.,  {Dekel} A.,  {Elmegreen} B.~G.,
  {Elmegreen} D.~M.,  {Teyssier} R.,  {Amram} P.,  {Daddi} E.,  {Duc} P.-A.,
  {Elbaz} D.,  {Epinat} B.,  {Gabor} J.~M.,  {Juneau} S.,  {Kraljic} K.,    {Le
  Floch'} E.,  2014, \apj, 780, 57

\bibitem[\protect\citeauthoryear{{Chien} \& {Barnes}}{{Chien} \&
  {Barnes}}{2010}]{Chien2010}
{Chien} L.-H.,  {Barnes} J.~E.,  2010, \mnras, 407, 43

\bibitem[\protect\citeauthoryear{{Cox}, {Jonsson}, {Somerville}, {Primack} \&
  {Dekel}}{{Cox} et~al.}{2008}]{Cox2008}
{Cox} T.~J.,  {Jonsson} P.,  {Somerville} R.~S.,  {Primack} J.~R.,    {Dekel}
  A.,  2008, \mnras, 384, 386

\bibitem[\protect\citeauthoryear{{Cullen}, {Alexander} \& {Clemens}}{{Cullen}
  et~al.}{2006}]{Cullen2006}
{Cullen} H.,  {Alexander} P.,    {Clemens} M.,  2006, \mnras, 366, 49

\bibitem[\protect\citeauthoryear{{Daddi}, {Elbaz}, {Walter}, {Bournaud},
  {Salmi}, {Carilli}, {Dannerbauer}, {Dickinson} \& {et al.}}{{Daddi}
  et~al.}{2010}]{Daddi2010b}
{Daddi} E.,  {Elbaz} D.,  {Walter} F.,  {Bournaud} F.,  {Salmi} F.,  {Carilli}
  C.,  {Dannerbauer} H.,  {Dickinson} M.,    {et al.} 2010, \apjl, 714, L118

\bibitem[\protect\citeauthoryear{{Dekel}, {Devor} \& {Hetzroni}}{{Dekel}
  et~al.}{2003}]{Dekel2003}
{Dekel} A.,  {Devor} J.,    {Hetzroni} G.,  2003, \mnras, 341, 326

\bibitem[\protect\citeauthoryear{{Di Matteo}, {Combes}, {Melchior} \&
  {Semelin}}{{Di Matteo} et~al.}{2007}]{DiMatteo2007}
{Di Matteo} P.,  {Combes} F.,  {Melchior} A.,    {Semelin} B.,  2007, \aap,
  468, 61

\bibitem[\protect\citeauthoryear{{Ellison}, {Patton}, {Simard} \&
  {McConnachie}}{{Ellison} et~al.}{2008}]{Ellison2008}
{Ellison} S.~L.,  {Patton} D.~R.,  {Simard} L.,    {McConnachie} A.~W.,  2008,
  \aj, 135, 1877

\bibitem[\protect\citeauthoryear{{Elmegreen}}{{Elmegreen}}{2011}]{Elmegreen2011}
{Elmegreen} B.~G.,  2011, \apj, 731, 61

\bibitem[\protect\citeauthoryear{{Elmegreen} \& {Scalo}}{{Elmegreen} \&
  {Scalo}}{2004}]{Elmegreen2004}
{Elmegreen} B.~G.,  {Scalo} J.,  2004, \araa, 42, 211

\bibitem[\protect\citeauthoryear{{Elmegreen}, {Elmegreen}, {Kaufman}, {Sheth},
  {Struck}, {Thomasson} \& {Brinks}}{{Elmegreen} et~al.}{2006}]{Elmegreen2006a}
{Elmegreen} D.~M.,  {Elmegreen} B.~G.,  {Kaufman} M.,  {Sheth} K.,  {Struck}
  C.,  {Thomasson} M.,    {Brinks} E.,  2006, \apj, 642, 158

\bibitem[\protect\citeauthoryear{{Elmegreen}, {Kaufman}, {Brinks}, {Elmegreen}
  \& {Sundin}}{{Elmegreen} et~al.}{1995}]{Elmegreen1995b}
{Elmegreen} D.~M.,  {Kaufman} M.,  {Brinks} E.,  {Elmegreen} B.~G.,    {Sundin}
  M.,  1995, \apj, 453, 100

\bibitem[\protect\citeauthoryear{{Emsellem} \& {van de Ven}}{{Emsellem} \& {van
  de Ven}}{2008}]{Emsellem2008}
{Emsellem} E.,  {van de Ven} G.,  2008, \apj, 674, 653

\bibitem[\protect\citeauthoryear{{Federrath}, {Klessen} \&
  {Schmidt}}{{Federrath} et~al.}{2008}]{Federrath2008}
{Federrath} C.,  {Klessen} R.~S.,    {Schmidt} W.,  2008, \apjl, 688, L79

\bibitem[\protect\citeauthoryear{{Federrath}, {Roman-Duval}, {Klessen},
  {Schmidt} \& {Mac Low}}{{Federrath} et~al.}{2010}]{Federrath2010}
{Federrath} C.,  {Roman-Duval} J.,  {Klessen} R.~S.,  {Schmidt} W.,    {Mac
  Low} M.-M.,  2010, \aap, 512, A81

\bibitem[\protect\citeauthoryear{{Federrath}, {Sur}, {Schleicher}, {Banerjee}
  \& {Klessen}}{{Federrath} et~al.}{2011}]{Federrath2011}
{Federrath} C.,  {Sur} S.,  {Schleicher} D.~R.~G.,  {Banerjee} R.,    {Klessen}
  R.~S.,  2011, \apj, 731, 62

\bibitem[\protect\citeauthoryear{{Frisch}}{{Frisch}}{1995}]{Frisch1995}
{Frisch} U.,  1995, {Turbulence. The legacy of A. N. Kolmogorov.}

\bibitem[\protect\citeauthoryear{{Gao} \& {Solomon}}{{Gao} \&
  {Solomon}}{2004}]{Gao2004}
{Gao} Y.,  {Solomon} P.~M.,  2004, \apjs, 152, 63

\bibitem[\protect\citeauthoryear{{Garc{\'{\i}}a-Burillo}, {Usero},
  {Alonso-Herrero}, {Graci{\'a}-Carpio}, {Pereira-Santaella}, {Colina},
  {Planesas} \& {Arribas}}{{Garc{\'{\i}}a-Burillo} et~al.}{2012}]{Garcia2012}
{Garc{\'{\i}}a-Burillo} S.,  {Usero} A.,  {Alonso-Herrero} A.,
  {Graci{\'a}-Carpio} J.,  {Pereira-Santaella} M.,  {Colina} L.,  {Planesas}
  P.,    {Arribas} S.,  2012, \aap, 539, A8

\bibitem[\protect\citeauthoryear{{Genzel}, {Tacconi}, {Gracia-Carpio},
  {Sternberg}, {Cooper}, {Shapiro}, {Bolatto} \& {et al.}}{{Genzel}
  et~al.}{2010}]{Genzel2010}
{Genzel} R.,  {Tacconi} L.~J.,  {Gracia-Carpio} J.,  {Sternberg} A.,  {Cooper}
  M.~C.,  {Shapiro} K.,  {Bolatto} A.,    {et al.} 2010, \mnras, 407, 2091

\bibitem[\protect\citeauthoryear{{Hancock}, {Smith}, {Struck}, {Giroux} \&
  {Hurlock}}{{Hancock} et~al.}{2009}]{Hancock2009}
{Hancock} M.,  {Smith} B.~J.,  {Struck} C.,  {Giroux} M.~L.,    {Hurlock} S.,
  2009, \aj, 137, 4643

\bibitem[\protect\citeauthoryear{{Hennebelle} \& {Falgarone}}{{Hennebelle} \&
  {Falgarone}}{2012}]{Hennebelle2012}
{Hennebelle} P.,  {Falgarone} E.,  2012, \aapr, 20, 55

\bibitem[\protect\citeauthoryear{{Hopkins}}{{Hopkins}}{2012}]{Hopkins2012b}
{Hopkins} P.~F.,  2012, \mnras, 423, 2037

\bibitem[\protect\citeauthoryear{{Hopkins}, {Somerville}, {Hernquist}, {Cox},
  {Robertson} \& {Li}}{{Hopkins} et~al.}{2006}]{Hopkins2006}
{Hopkins} P.~F.,  {Somerville} R.~S.,  {Hernquist} L.,  {Cox} T.~J.,
  {Robertson} B.,    {Li} Y.,  2006, \apj, 652, 864

\bibitem[\protect\citeauthoryear{{Irwin}}{{Irwin}}{1994}]{Irwin1994}
{Irwin} J.~A.,  1994, \apj, 429, 618

\bibitem[\protect\citeauthoryear{{Jog}}{{Jog}}{2013a}]{Jog2013a}
{Jog} C.~J.,  2013a, \mnras, 434, L56

\bibitem[\protect\citeauthoryear{{Jog}}{{Jog}}{2013b}]{Jog2013b}
{Jog} C.~J.,  2013b, ArXiv e-prints

\bibitem[\protect\citeauthoryear{{Jog} \& {Solomon}}{{Jog} \&
  {Solomon}}{1992}]{Jog1992}
{Jog} C.~J.,  {Solomon} P.~M.,  1992, \apj, 387, 152

\bibitem[\protect\citeauthoryear{{Karl}, {Naab}, {Johansson}, {Kotarba},
  {Boily}, {Renaud} \& {Theis}}{{Karl} et~al.}{2010}]{Karl2010}
{Karl} S.~J.,  {Naab} T.,  {Johansson} P.~H.,  {Kotarba} H.,  {Boily} C.~M.,
  {Renaud} F.,    {Theis} C.,  2010, \apjl, 715, L88

\bibitem[\protect\citeauthoryear{{Keel}, {Kennicutt} Jr., {Hummel} \& {van der
  Hulst}}{{Keel} et~al.}{1985}]{Keel1985}
{Keel} W.~C.,  {Kennicutt} Jr. R.~C.,  {Hummel} E.,    {van der Hulst} J.~M.,
  1985, \aj, 90, 708

\bibitem[\protect\citeauthoryear{{Khochfar} \& {Burkert}}{{Khochfar} \&
  {Burkert}}{2006}]{Khochfar2006}
{Khochfar} S.,  {Burkert} A.,  2006, \aap, 445, 403

\bibitem[\protect\citeauthoryear{{Kraljic}, {Renaud}, {Bournaud}, {Combes},
  {Elmegreen}, {Emsellem} \& {Teyssier}}{{Kraljic} et~al.}{2014}]{Kraljic2014}
{Kraljic} K.,  {Renaud} F.,  {Bournaud} F.,  {Combes} F.,  {Elmegreen} B.,
  {Emsellem} E.,    {Teyssier} R.,  2014, \apj, 784, 112

\bibitem[\protect\citeauthoryear{{Kritsuk}, {Norman}, {Padoan} \&
  {Wagner}}{{Kritsuk} et~al.}{2007}]{Kritsuk2007}
{Kritsuk} A.~G.,  {Norman} M.~L.,  {Padoan} P.,    {Wagner} R.,  2007, \apj,
  665, 416

\bibitem[\protect\citeauthoryear{{Krumholz}, {Dekel} \& {McKee}}{{Krumholz}
  et~al.}{2012}]{Krumholz2012}
{Krumholz} M.~R.,  {Dekel} A.,    {McKee} C.~F.,  2012, \apj, 745, 69

\bibitem[\protect\citeauthoryear{{Krumholz} \& {Tan}}{{Krumholz} \&
  {Tan}}{2007}]{Krumholz2007a}
{Krumholz} M.~R.,  {Tan} J.~C.,  2007, \apj, 654, 304

\bibitem[\protect\citeauthoryear{{Longmore}, {Bally}, {Testi}, {Purcell},
  {Walsh}, {Bressert}, {Pestalozzi}, {Molinari} \& {et al.}}{{Longmore}
  et~al.}{2013}]{Longmore2013}
{Longmore} S.~N.,  {Bally} J.,  {Testi} L.,  {Purcell} C.~R.,  {Walsh} A.~J.,
  {Bressert} E.,  {Pestalozzi} M.,  {Molinari} S.,    {et al.} 2013, \mnras,
  429, 987

\bibitem[\protect\citeauthoryear{{Mac Low} \& {Klessen}}{{Mac Low} \&
  {Klessen}}{2004}]{MacLow2004}
{Mac Low} M.-M.,  {Klessen} R.~S.,  2004, Reviews of Modern Physics, 76, 125

\bibitem[\protect\citeauthoryear{{Molina}, {Glover}, {Federrath} \&
  {Klessen}}{{Molina} et~al.}{2012}]{Molina2012}
{Molina} F.~Z.,  {Glover} S.~C.~O.,  {Federrath} C.,    {Klessen} R.~S.,  2012,
  \mnras, 423, 2680

\bibitem[\protect\citeauthoryear{{Nordlund} \& {Padoan}}{{Nordlund} \&
  {Padoan}}{1999}]{Nordlund1999}
{Nordlund} {\AA}.~K.,  {Padoan} P.,  1999, in Proc. 2nd Guillermo Haro
  Conference on Interstellar Turbulence the Density PDFs of Supersonic Random
  Flows, ed. J. Franco \& A. Carraminana (Cambridge: Cambridge Univ. Press)
  p.~218

\bibitem[\protect\citeauthoryear{{Powell}, {Bournaud}, {Chapon} \&
  {Teyssier}}{{Powell} et~al.}{2013}]{Powell2013}
{Powell} L.~C.,  {Bournaud} F.,  {Chapon} D.,    {Teyssier} R.,  2013, \mnras,
  434, 1028

\bibitem[\protect\citeauthoryear{{Renaud}, {Boily}, {Fleck}, {Naab} \&
  {Theis}}{{Renaud} et~al.}{2008}]{Renaud2008}
{Renaud} F.,  {Boily} C.~M.,  {Fleck} J.-J.,  {Naab} T.,    {Theis} C.,  2008,
  \mnras, 391, L98

\bibitem[\protect\citeauthoryear{{Renaud}, {Boily}, {Naab} \& {Theis}}{{Renaud}
  et~al.}{2009}]{Renaud2009}
{Renaud} F.,  {Boily} C.~M.,  {Naab} T.,    {Theis} C.,  2009, \apj, 706, 67

\bibitem[\protect\citeauthoryear{{Renaud}, {Kraljic} \& {Bournaud}}{{Renaud}
  et~al.}{2012}]{Renaud2012}
{Renaud} F.,  {Kraljic} K.,    {Bournaud} F.,  2012, \apjl, 760, L16

\bibitem[\protect\citeauthoryear{{Robertson}, {Bullock}, {Cox}, {Di Matteo},
  {Hernquist}, {Springel} \& {Yoshida}}{{Robertson}
  et~al.}{2006}]{Robertson2006}
{Robertson} B.,  {Bullock} J.~S.,  {Cox} T.~J.,  {Di Matteo} T.,  {Hernquist}
  L.,  {Springel} V.,    {Yoshida} N.,  2006, \apj, 645, 986

\bibitem[\protect\citeauthoryear{{Robertson} \& {Kravtsov}}{{Robertson} \&
  {Kravtsov}}{2008}]{Robertson2008}
{Robertson} B.~E.,  {Kravtsov} A.~V.,  2008, \apj, 680, 1083

\bibitem[\protect\citeauthoryear{{Saintonge}, {Tacconi}, {Fabello}, {Wang},
  {Catinella}, {Genzel}, {Graci{\'a}-Carpio}, {Kramer} \& {et al.}}{{Saintonge}
  et~al.}{2012}]{Saintonge2012}
{Saintonge} A.,  {Tacconi} L.~J.,  {Fabello} S.,  {Wang} J.,  {Catinella} B.,
  {Genzel} R.,  {Graci{\'a}-Carpio} J.,  {Kramer} C.,    {et al.} 2012, \apj,
  758, 73

\bibitem[\protect\citeauthoryear{{Sanders} \& {Mirabel}}{{Sanders} \&
  {Mirabel}}{1996}]{Sanders1996}
{Sanders} D.~B.,  {Mirabel} I.~F.,  1996, \araa, 34, 749

\bibitem[\protect\citeauthoryear{{Saury}, {Miville-Desch{\^e}nes},
  {Hennebelle}, {Audit} \& {Schmidt}}{{Saury} et~al.}{2013}]{Saury2014}
{Saury} E.,  {Miville-Desch{\^e}nes} M.-A.,  {Hennebelle} P.,  {Audit} E.,
  {Schmidt} W.,  2013, ArXiv e-prints

\bibitem[\protect\citeauthoryear{{Seifried}, {Schmidt} \&
  {Niemeyer}}{{Seifried} et~al.}{2011}]{Seifried2011}
{Seifried} D.,  {Schmidt} W.,    {Niemeyer} J.~C.,  2011, \aap, 526, A14

\bibitem[\protect\citeauthoryear{{Smith}, {Struck}, {Hancock}, {Giroux},
  {Appleton}, {Charmandaris}, {Reach}, {Hurlock} \& {Hwang}}{{Smith}
  et~al.}{2008}]{Smith2008}
{Smith} B.~J.,  {Struck} C.,  {Hancock} M.,  {Giroux} M.~L.,  {Appleton} P.~N.,
   {Charmandaris} V.,  {Reach} W.,  {Hurlock} S.,    {Hwang} J.-S.,  2008, \aj,
  135, 2406

\bibitem[\protect\citeauthoryear{{Tasker} \& {Bryan}}{{Tasker} \&
  {Bryan}}{2008}]{Tasker2008}
{Tasker} E.~J.,  {Bryan} G.~L.,  2008, \apj, 673, 810

\bibitem[\protect\citeauthoryear{{Teyssier}}{{Teyssier}}{2002}]{Teyssier2002}
{Teyssier} R.,  2002, \aap, 385, 337

\bibitem[\protect\citeauthoryear{{Teyssier}, {Chapon} \& {Bournaud}}{{Teyssier}
  et~al.}{2010}]{Teyssier2010}
{Teyssier} R.,  {Chapon} D.,    {Bournaud} F.,  2010, \apjl, 720, L149

\bibitem[\protect\citeauthoryear{{Ueda}, {Iono}, {Petitpas}, {Yun}, {Ho},
  {Kawabe}, {Mao}, {Mart{\'{\i}}n} \& {et al.}}{{Ueda} et~al.}{2012}]{Ueda2012}
{Ueda} J.,  {Iono} D.,  {Petitpas} G.,  {Yun} M.~S.,  {Ho} P.~T.~P.,  {Kawabe}
  R.,  {Mao} R.-Q.,  {Mart{\'{\i}}n} S.,    {et al.} 2012, \apj, 745, 65

\bibitem[\protect\citeauthoryear{{Valluri}}{{Valluri}}{1993}]{Valluri1993}
{Valluri} M.,  1993, \apj, 408, 57

\bibitem[\protect\citeauthoryear{{Vazquez-Semadeni}}{{Vazquez-Semadeni}}{1994}]{Vazques1994}
{Vazquez-Semadeni} E.,  1994, \apj, 423, 681

\bibitem[\protect\citeauthoryear{{Wada} \& {Norman}}{{Wada} \&
  {Norman}}{2001}]{Wada2001}
{Wada} K.,  {Norman} C.~A.,  2001, \apj, 547, 172

\bibitem[\protect\citeauthoryear{{Wang}, {Fazio}, {Ashby}, {Huang}, {Pahre},
  {Smith}, {Willner}, {Forrest} \& {et al.}}{{Wang} et~al.}{2004}]{Wang2004}
{Wang} Z.,  {Fazio} G.~G.,  {Ashby} M.~L.~N.,  {Huang} J.-S.,  {Pahre} M.~A.,
  {Smith} H.~A.,  {Willner} S.~P.,  {Forrest} W.~J.,    {et al.} 2004, \apjs,
  154, 193

\end{thebibliography}

\end{document}